\documentclass[a4paper,10pt,twoside]{article}
\usepackage{amsmath}
\usepackage{amssymb}
\usepackage{amsthm}
\usepackage{graphicx}

\theoremstyle{definition}

\theoremstyle{remark}

\newcommand{\bk}{{\bf k}}

\newcommand{\bR}{{\bf R}}

\newcommand{\bB}{{\bf B}}

\newcommand{\bx}{{\bf x}}
\newcommand{\bz}{{\bf z}}
\newcommand{\by}{{\bf y}}
\newcommand{\bp}{{\bf p}}
\newcommand{\bC}{{\bf C}}
\newcommand{\bq}{{\bf q}}

\newcommand{\hk}{{\hat{\bf k}}}

\newcommand{\hx}{{\hat{\bf x}}}
\newcommand{\hy}{{\hat{\bf y}}}

\newcommand{\hp}{{\hat{\bf p}}}
\newcommand{\hq}{{\hat{\bf q}}}

\newcommand{\hz}{{\hat{\bf z}}}

\newcommand{\be}{\begin{equation}}
\newcommand{\ee}{\end{equation}}
\newcommand{\bay}{\begin{eqnarray}}
\newcommand{\eay}{\end{eqnarray}}
 \author{V.~S.~Buslaev, S.~B.~Levin}
 \title{ The system of three three-dimensional charged quantum particles: asymptotic behavior of the eigenfunctions of the continuous spectrum at infinity}
\begin{document}
 \addtolength{\hoffset}{-3.1cm}
  \addtolength{\voffset}{-3cm}

\maketitle

\begin{center}
{Department of Mathematical and Computational Physics,\\
St-Petersburg State University, Russia}
\end{center}

\begin{center}
\end{center}

{\bf{Abstract}}. To our knowledge there are no complete results expressed in terms of eigenfunctions (even not strictly proved mathematically) related to the system of three or more charged quantum particles.
For the system of the three such identical particles we suggest
the asymptotic formula describing the behavior of eigenfunctions at infinity in configuration space.

{\bf{Introduction}}. The principal questions of the scattering theory in the system of two, three and more quantum particles, considered in terms of eigenfunctions (EF), have been studied relatively completely \cite{LD1} but not exhaustively. In this field of problems there is a set of important systems which so far remain unexplored. Here we mean the systems of three and more particles interacting by pair coulomb potentials. There are two widely practised (for computer calculations) approaches to such problems.
The first one consists in the regularization of the coulomb potential by its substitution by for example
the Yukawa potential. Until now, however, there was no possibility to estimate the effect of such a regularization, as no reliable (even not strictly proved) results for many-body systems with the pair coulomb interactions were known.

The second approach consists in the application of the so called BBK-approximation to the continuum spectrum EF, refer \cite{BBK}. We will describe it in detail later. In a few words, BBK-approximation can be presented by the explicit formula which has a discrepancy in the equation, decreasing at infinity quickly enough. The problem is that on some specific multi-dimensional directions at infinity, which we will call the screens where the particles approach each other in pairs, the discrepancy loses a satisfactory pace of decrease. Our result consists in the modification of BBK-approximation on the screens, so that the modified formula allows a satisfactory description of the EF asymptotics at all directions at infinity. Thus, we can say that we have succeeded for the first time in deriving the asymptotic behavior of EF. To estimate the importance of this result, note that when considering the EF even for the quickly decreasing potentials, we always start the constructions with the derivation of the asymptotics of these EF (or, which is almost equivalent) with the derivation of the corresponding integral equations.

{\bf{Weak asymptotics}}. We consider the system of three three-dimensional particles of equal masses,
interacting by identical pair coulomb potentials. The assumption about equality of masses and potentials
is introduced only for the simplicity of the presentation. These restrictions can easily removed. It is quite obvious that the system has a pure continuous and infinitely-tuple degenerate spectrum spanning the half-axis $[0, \infty)$. The traditional way of the identification of the EF is connected with the
description of its possible behavior at infinity in the configuration space of the system. For the quickly
decreasing at infinity pair potentials such an identification of the EF consists for example in the separation (as the main order) of the plane wave with the fixed wave vector
from its asymptotics. For the the case of the coulomb potentials the plane wave can not be considered
as the main order of the asymptotics of the EF in conventional sense.

Even for the case of one particle scattering on the coulomb potential  $v,\ v(\bx) = \frac{\alpha}{x}$, $\bx \in \bf{R}^3$, $\hx = \frac{\bx}{x}$, such a simple separation of the main order term in the asymptotic behavior of the solution is impossible. The simplest description can be done in terms of the so called
weak asymptotics \cite{VSB}. It can be presented by the formula \cite{MF}:
\be
\psi(\bx,\bk)\ \sim \frac{2\pi i}{kx}\left(\delta(\hx,-\hk)e^{-ikx+i\eta\ln x}-s_c(\hx,\bk)e^{ikx-i\eta\ln x}\right), \quad \eta = \frac{\alpha}{2|\bk|},\ \ \ |\bx| \to \infty.
\label{eq1}
\ee
Here the coefficient $s_c$, in fact the scattering matrix, in the problem set up process remains
undefined and is determined only by the complete solution of the problem. The main feature
of this asymptotic description is that we consider it in terms of distributions relatively to the $\hx $.
In spite of the weakened character of the condition, it successfully separates the unique solution, standard EF.

To make it easier for the reader to compare this formula with the classical assymptotic description of the EFs, note that in the weak sense
$e^{i<\bx,\bk>}\sim\frac{2\pi i}{kx}\left(
\delta(\hx,-\hk)e^{-ikx}- \delta(\hx,\bk)e^{ikx}\right).$

Turn to the system of three particles, fix the motion of the center of mass and will describe the internal degrees of freedom in the terms of the variable $\bz$ (details follow below).
Then the assymptotics of EF $ \Psi(\bz,\bq)$, $ \bq$ - is a wave vector, which can be considered as a weak alternative of the plain wave for the coulomb potentials, characterized by the formula
$$
\Psi(\bz,\bq)  \sim \frac12\left(\frac{4\pi i}{qz}\right)^{5/2}\delta(\hq,-\hz)e^{-iqz+\mathop{\sum}\limits_{j=1}^3\eta_j\ln z}-
\frac12\left(\frac{4\pi i}{qz}\right)^{5/2}S_c(\bq,\hz)e^{iqz-\mathop{\sum}\limits_{j=1}^3\eta_j\ln z},
$$

The system of descriptions for the vectors reproduces here the descriptions used in the formula (\ref{eq1}), only with the difference that $\bz \in \Gamma, \dim \Gamma = 6$.
Here again $S_c$ - is a function indefinite when setting up a problem, which as a final result must turn into a scattering matrix.

We definitely believe that such a characterization of the EF, just as in the case of quickly decreasing potentials, determines these functions in the unique way.

{\it{ The main result of the work is that, starting from a simple asymptotic characterization of EF in the weak sense, we get their asymptotic characterization in the traditional, uniform, pointwise sense}}.

The possibility to use an explicit weak asymptotics for the description of the solution allows basically to determine a solution. On the other hand in the work \cite{BL1} it is shown how one can use uniform asymptotic formulas for the numerical description of the EF. Besides, the result achieved offers quite a convincing hope for its possible development and derivation of the traditional complete proof that there exists a unique solution with the asymptotic behavior achieved here. It is not simple and in the present work we do not go beyond the heuristic construction of the asymptotic formulas for EF.

{\bf{Description of the model }}. Let us consider a more detailed description of the model. The initial configurational space of the system is $\bR^9$. Having stopped the motion of the center of the masses, we arrive at the system on the configurational space $\Gamma=\{\bz:\, \bz\in\bR^9,\,\bz=\{\bz_1,\bz_2,\bz_3\},\ \bz_1+\bz_2+\bz_3=0\}.$ On $\Gamma$ there is a scalar product $\langle\bz,\bz'\rangle$, induced by the scalar product on $\bR^9$. The system at $\Gamma$ is described by the equation
$H\Psi=\lambda\Psi,\ \ \ \Psi=\Psi(\bz)\in\bC,\ \ \bz\in\Gamma,\ \ \ \
H=-\Delta_\bz+V(\bz), \quad V(\bz) = v(\bx_1)+v(\bx_2)+v(\bx_3),\ \ \ \bx_j\in\bR^3.$
Here $\Delta_\bz$ -- is the Laplace operator at $\Gamma$, $\ \bx_1=\frac{1}{\sqrt{2}}(\bz_3-\bz_2), \bx_2=\frac{1}{\sqrt{2}}(\bz_1-\bz_3), \bx_3=\frac{1}{\sqrt{2}}(\bz_2-\bz_1)$ . It is clear that $\bx_1+\bx_2+\bx_3=0$.
Let us introduce also $\by_j=\sqrt{\frac{3}{2}}\bz_j$. It is easy to verify that at $\Gamma$ $\by_1+\by_2+\by_3=0,$
$\bz^2=\langle\bz,\bz\rangle=\langle\bx_j,\bx_j\rangle+\langle\by_j,\by_j\rangle,\ \ \ j=1,2,3\ \ \ , \quad \Delta_\bz = \Delta_\bx + \Delta_\by.$

Together with $\bz\in\Gamma,\ \bx,\by\in\bR^3$ we will consider the dual variables, momenta $\bq \in \Gamma,\bk,\bp \in \bR^3$.
We will assume that
$v(\bx)=\frac{\alpha}{x},\ \ \alpha>0,$
though a generalization is possible for the case $v(\bx)= \frac{\alpha}{x} + w(\bx),\ xw(\bx) \to 0,\ x \to \infty.$

{\bf{BBK - approximation}}. The plain wave at asymptotic description of the function $\Psi(\bz,\bq)$ beyond the vicinities of the screens $\sigma_{j} = \{\bz \in \Gamma,\ \bx_j = 0\}$, $\ j = 1,2,3,$ must be replaced by the BBK-approximation $\Psi^{BBK}(\bz,\bq)$. This approximation was studied in \cite{BBK}, refer also to \cite{MF}, though it had been used also earlier in \cite{z1,z2}. It acquires the form:
\be
\Psi^{BBK}(\bz,\bq)\sim N_0 e^{i\langle\bz,\bq\rangle}D(\bx_1,\bk_1)D(\bx_2,\bk_2)D(\bx_3,\bk_3).
\label{bbk0}
\ee

Here $D(\bx,\bk) = \Phi(-i\eta, 1, ixk - i<\bx,\bk>)$, $\ \bx,\bk \in {\bR}^3,$ $\eta = \frac{\alpha}{2k}$, $\Phi$ - is a confluent hypergeometric function, refer \cite{GR}.The constant
$N_0=\mathop{\prod}\limits_{j=1}^3N_c^{(j)}$ is a product of the normalization constants of the three two-body scattering states
$N_c^{(j)}=(2\pi)^{-\frac32}e^{-\frac{\pi\eta_j}{2}}\Gamma(1+i\eta_j)$.

It is noteworthy that the function
\be
\psi_c(\bx,\bk) = N_c e^{i<\bx,\bk>}D(\bx, \bk)
\label{pair}
\ee
is the solution of the one particle scattering problem on the coulomb potential.

The discrepancy $BBK$ of the approximation $Q[\Psi^{BBK}]$ can be easily calculated:
$$
Q[\Psi^{BBK}] = -k_2k_3<\hk_2-\hx_2,\hk_3-\hx_3>\Phi_{1}\Phi_2^{'}\Phi_3^{'}  -k_3k_1<\hk_1-\hx_1,\hk_3-\hx_3>\Phi_{1}^{'}\Phi_2\Phi_3^{'} - k_1k_2<\hk_2-\hx_2,\hk_1-\hx_1>
\Phi_{1}^{'}\Phi_2^{'}\Phi_3,
$$
where $'$ defines the derivative by the last argument of the function $\Phi$. With $z \to \infty $ beyond some vicinities of the screens the discrepancy decreases quicker than the coulomb potential. Though the BBK-approximation is continuous up to the screens, its discrepancy on the screens decreases not quicker than the coulomb potential. Because of that the BBK-approximation cannot be used in the vicinities of the screens.

{\bf{Separation of variables}}. In the vicinity of each screen the Schredinger equation allows an essential simplification. For example, the complete potential $V(\bz)= v(\bx_1) + v(\bx_2) + v(\bx_3)$ in the vicinity of the screen $\sigma_1$ is simplified because of the formulas
$\bx_2 = - \frac{\sqrt{3}}{2}\by -\frac{1}{2}\bx,\ \bx_3 = \frac{\sqrt{3}}{2}\by -\frac{1}{2}\bx,\
\bx = \bx_1,\ \by = \by_1.$

In the vicinity $\sigma_1$ $\ y>>1,\  y >> x $,
therefore the formula
$V \sim v(x) + v_m(y),\ \ v_m = \frac{4\alpha/\sqrt{3}}{y}$
gives a good approximation to the potential.
An equation with such a potential allows a separation of variables
$-\Delta_{\bz}\chi + (v(x) + v_m(y))\chi = \lambda \chi.$

Since we are interested in restricted solutions, for $\chi$ naturally arises the following representation
$$
\chi(\bx,\by,\bk,\bp)=\int\psi_c(\bx,\bk')\psi_m(\by,\bp')\delta({k'}^2+{p'}^2-E)R(\bq,\bq')d\bk' d\bp',\ \ \ \
\bq=(\bk,\bp),\ \ \bq'=(\bk',\bp').
$$

Here $\psi_c$ -- is the solution of the scattering problem for the potential $v$, while $\psi_m$ -- is the solution of the scattering problem for the potential
$v_m$.

We can formulate now the goal of the present work more specifically. We are going to give an explicit description of the leading order of the asymptotic $\Psi$ in the uniform topology relative to the angle variable, $\hz$.
Precisely this problem is solved here on the heuristic level. Note that having this result, one can change the set-up of the problem (we assume that the vector $\bq$ is beyond some small vicinities of the three screens) and start finding a solution, characterizing it by the uniform asymptotics.

{\bf{BBK--approximation in the vicinity of the screen}}.BBK--approximation in the vicinity, for example, of the screen $\sigma_1$ but, however, in the domain where the discrepancy of this approximation decreases still quicker than the potential, naturally separates into a product \be
\Psi_{BBK}(\bz,\bq)=\psi_c(\bx_1,\bk_1)\Psi_1(\bz,\bq),\ \ \ \Psi_1(\bz,\bq) = N_0^{(23)}e^{i\langle\by_1,\bp_1\rangle}
D(\bx_2,\bk_2)D(\bx_3,\bk_3),\ \ \ \ N_0^{(23)}=N_c^{(2)}N_c^{(3)}.
\label{bi0}
\ee

In the vicinity of the screen the variables $\bx_1,\ \by_1$ asymptotically have a different order, $y_1\gg x_1$. Let us compute the weak asymptotics in these assumptions
$\Psi_1(\bz,\bq)$ for $y_1\rightarrow\infty$
\be
\Psi_1(\bz,\bq)=
\delta(\hat{\bp_1},-\hat{\by_1})B_0(\bq)\frac{2\pi}{iy_1p_1}e^{-iy_1p_1+i\omega\ln y_1}
e^{i\eta_2\ln\left[Z_2^- + \frac12\frac{x_1}{y_1}V_2^-\right]}
e^{i\eta_3\ln\left[Z_3^- + \frac12\frac{x_1}{y_1}V_3^-\right]}-
\label{bi11}
\ee
$$
-
\delta(\hat{\bp_1},\hat{\by_1})B_0(\bq)\frac{2\pi}{iy_1p_1}e^{iy_1p_1+i\omega\ln y_1}
e^{i\eta_2\ln\left[Z_2^+ + \frac12\frac{x_1}{y_1}V_2^+\right]}
e^{i\eta_3\ln\left[Z_3^+ + \frac12\frac{x_1}{y_1}V_3^+\right]}.
$$

Here we have used the following notations
$Z_2^\pm=\frac{\sqrt{3}}{2}(1\pm \langle\hat{\bp_1},\hat{\bk}_2\rangle),\ \ \ \ \ \
Z_3^\pm=\frac{\sqrt{3}}{2}(1\mp \langle\hat{\bp_1},\hat{\bk}_3\rangle),\ \ \ \ \ \
V_2^\pm=\langle\hx_1,\hk_2\pm\hp_1\rangle,\ \ \ \ \ \
V_3^\pm=\langle\hx_1,\hk_3\mp\hp_1\rangle,$
$\omega=\frac{\alpha}{2k_2}+\frac{\alpha}{2k_3},\ \ \ \ \ \ \ \
B_0(\bq)=B_0(\bq)=-(2\pi)^{-3}k_2^{i\eta_2}k_3^{i\eta_3}.$

Coefficients $Z_{2(3)}^+,\ V_{2(3)}^+$ differ from the coefficients $Z_{2(3)}^-,\ V_{2(3)}^-$
by the change of the vector $\hp_1$ for $-\hp_1$.

{\bf{Computation of the coefficient $R$}}. We will certainly need the asymptotic behavior $\chi$ for
$y\rightarrow\infty$. Let us substitute the weak asymptotics $\psi_m$ into the integral $\chi$:
\be
\chi\sim \int\psi_c(\bx,\bk)\frac{2\pi i}{p y}\left(\delta(\hy,-\hp)e^{-ipy+i\eta_m\ln y}-S_m(\hy,\bp)
e^{ipy-i\eta_m\ln y}\right)\delta({k'}^2+{p'}^2-E)R(\bq,\bq')d\bk' d\bp'.
\label{as1}
\ee

Here $S_m$ -- is the coulomb scattering matrix corresponding to the potential $v_m$.

For the further asymptotic simplification of the integral $\chi$ for $y\rightarrow\infty$ we need the information on the structure of the coefficient $R$. It can be obtained from the comparison of the weak asymptotics on the variable $\by$ (the direct pointwise comparison turns out to be a significantly more complicated problem) of the integral (\ref{as1}) and
of the BBK--approximation (\ref{bi0}-\ref{bi11}) in the domain, where the discrepancies of both approximations decrease quicker than the potential.
It is a rather cumbersome computation that we are not able to demonstrate in the present paper.
But it is the most crucial part of the work. The main idea of the computation is, however, quite natural.
The representation of the $BBK-$ approximation at some distance from the screen by the integral $\chi$ is, in fact, the spectral resolution with the respect of the eigenfunctions  $\psi_m$ , and therefore, the resolution coefficient, i.e. the coefficient  $R$, can be found explicitly.

The result is:
\be
R(\bq,\bq')=\frac{1}{kpk'p'}A_{in}(\bq)\frac{\delta(\hat{\bp},\hat{\bp'})}{(p'-p-i0)^{1+ia}}
\delta\left(\hat{\bk}',\frac{\hat{\bk}+(p'-p)\bB_{in}}{|\hat{\bk}+(p'-p)\bB_{in}|}\right)+
\label{kern01}
\ee
$$
+\frac{1}{kpk'p'}A_{out}(\bq)\frac{G(\hat{\bp'},{\bp})}{(p'-p+i0)^{1+ib}}
\delta\left(\hat{\bk}',\frac{\hat{\bk}+(p'-p)\bB_{out}}{|\hat{\bk}+(p'-p)\bB_{out}|}\right).
$$

Here the kernel $G=S_m^{-1}$ satisfies the equation $\int S_m(\hy,p\hp')G(\hp',\bp)d\hp'=\delta(\hy,\hp)$,
$\ \ a=\omega-\frac{2\alpha}{\sqrt{3}p},\ \ \ \ \ b=\omega+\frac{2\alpha}{\sqrt{3}p},\ \ \ \ \
\omega=\frac{\alpha}{2k_2}+\frac{\alpha}{2k_3},$

The coefficient $A_{in}$ and the vector $\bB_{in}$ can be found from the equations
$\ A_{in}=-\frac{k}{\pi i}\Gamma(1-ia)e^{\frac{\pi a}{2}}B_0^{in},\ \ \ \ \ \ \
\bB_{in}=\frac{p}{k^2}\hk-\frac{1}{ak}\Omega_{in},\ \ \ \langle\bB_{in},\hk\rangle=0.$

Correspondingly, the coefficient  $A_{out}$ and the vector  $\bB_{out}$ can be found from the equations
$\ A_{out}=\frac{k}{\pi i}\Gamma(1-ib)e^{-\frac{\pi b}{2}}B_0^{out},\ \ \ \ \ \ \
\bB_{out}=\frac{p}{k^2}\hk-\frac{1}{bk}\Omega_{out},\ \ \ \langle\bB_{out},\hk\rangle=0.$

We used here the notations
$$
B_0^{in}(\bq)=(2\pi)^{-3}
\left[\frac{\sqrt{3}}{2}(1-\langle\hat{\bp},\hat{\bk}_2\rangle)\right]^{i\eta_2}
\left[\frac{\sqrt{3}}{2}(1+\langle\hat{\bp},\hat{\bk}_3\rangle)\right]^{i\eta_3}k_2^{i\eta_2}k_3^{i\eta_3},
$$
$$
B_0^{out}(\bq)=(2\pi)^{-3}
\left[\frac{\sqrt{3}}{2}(1+\langle\hat{\bp},\hat{\bk}_2\rangle)\right]^{i\eta_2}
\left[\frac{\sqrt{3}}{2}(1-\langle\hat{\bp},\hat{\bk}_3\rangle)\right]^{i\eta_3}k_2^{i\eta_2}k_3^{i\eta_3},
$$
$$
\Omega_{in}= \frac{1}{\sqrt{3}}\left(\eta_2\frac{\hat{\bk}_2-
\hat{\bp}}{1-\langle\hat{\bp},\hat{\bk}_2\rangle}+
\eta_3\frac{\hat{\bk}_3+
\hat{\bp}}{1+\langle\hat{\bp},\hat{\bk}_3\rangle}\right),\ \ \ \ \ \
\Omega_{out}= \frac{1}{\sqrt{3}}\left(\eta_2\frac{\hat{\bk}_2+
\hat{\bp}}{1+\langle\hat{\bp},\hat{\bk}_2\rangle}+
\eta_3\frac{\hat{\bk}_3-
\hat{\bp}}{1-\langle\hat{\bp},\hat{\bk}_3\rangle}\right).
$$

With such a choice of  $R$ the asymptotic in the weak sense formula  (\ref{as1}) for $\chi$ and the asymptotic formula   (\ref{bi0})-(\ref{bi11})
for $\Psi^{BBK}_{j}$ for $y\rightarrow\infty$ coincide in the leading order on the intersection of the domains
$$
V_{0j}=\{\bz:\ x_j < y_j^{\nu}, 0 < \nu < 1\},\ \quad
V_{1j}=\{\bz:\ x_j > y_j^{\mu}, 1/2 < \mu < \nu\}.
$$

The next step is the derivation of the weak asymptotics of the integral $\chi$ (\ref{as1}) with $y\gg 1$ and arbitrary $\ x$, $\ x\ll y$ and the reconstruction of the pointwise asymptotics by the weak asymptotics in the vicinity of the screens:
\be
\chi(\bz,\bq)\sim N_0 e^{i\langle\bz,\bq\rangle}D(\bx_1,\bk_1)D(\tilde{\bx}_2,\bk_2)D(\tilde{\bx}_3,\bk_3).
\label{bbk01}
\ee

This expression coincides with the expression (\ref{as1}) for $\Psi^{BBK}$ with the substitution
$$
\bx_2=-\frac{\sqrt{3}}{2}\by-\frac12\bx\ \ \ \rightarrow\ \ \
\tilde{\bx}_2=-\frac{\sqrt{3}}{2}\by+i\frac12\frac{\nabla_\bk\psi_c(\bx,\bk)}{\psi_c(\bx,\bk)},
$$
 $$
\bx_3=\frac{\sqrt{3}}{2}\by-\frac12\bx\ \ \ \rightarrow\ \ \
\tilde{\bx}_3=\frac{\sqrt{3}}{2}\by+i\frac12\frac{\nabla_\bk\psi_c(\bx,\bk)}{\psi_c(\bx,\bk)},
$$
where the functions $\psi_c(\bx,\bk)$ are defined in the equation (\ref{pair}).

This result is the central one. Note that at large $x$ the expressions $\tilde{\bx}_{2(3)}$
turn into the ordinary expressions ${\bx}_{2(3)}$, while the integral (\ref{bbk01}) turns into $\Psi^{BBK}$
(\ref{bbk0}). Note as well, that beyond the vicinities of the pair directions of the forward scattering
 $\langle\hat{\bx}_j,\hat{\bk}_j\rangle=1,\ \ j=1,2,3$ , our result corresponds to the result of the work
\cite{am}. The uniform in angles description of the leading order of the scattering problem solution asymptotics has been obtained
for the first time.

{\bf{Formulating of the result}}.
The sets $V_0=\Gamma\backslash\mathop{\bigcup}\limits_{j=1}^3 V_{0j}$ and
$V_{1}=\mathop{\bigcup}\limits_{j=1}^3 V_{0j}$ cover $\Gamma$: $\Gamma =
V_{0} \cup V_{1}$. Consider the separation of the unit $1 = \zeta _{0}(x,y) + \sum_{j=1}^3\zeta_{0j}(x,y)$, subordinated to this covering. Let us assume that there, where the functions $\zeta _{0j},\ \zeta _{0}$ differ from the constants, they depend on the ratio $\rho = lnx_j / lny_j$.

Let us form the expression
$$
\Psi^{as}_j = \zeta_{0j} \chi_j.
$$

Let us also define on $\Gamma$ the function
$$
\Psi^{as} = \sum_{j=1}^3 \Psi^{as}_j + \zeta_{0}\Psi^{BBK}.
$$

We believe in the correctness of the following main assertion:
{\it Function $\Psi^{as}$ correctly describes the asymptotic behavior of the solution $\Psi$ in the leading order in the uniform topology.}
One can also verify that the discrepancy $Q[\Psi^{as}] = - \Delta_{\bf{z}} \Psi^{as} + V \Psi^{as} - \lambda \Psi^{as}$ decreases at $z \to \infty$ quicker than the coulomb potential.

\end{document}